\documentclass[twocolumn]{aastex631}

\usepackage{xspace}	    
\usepackage{lineno}
\usepackage{multirow}   

\begin{document}
\newcommand{\vdag}{(v)^\dagger}
\newcommand{\myemail}{skywalker@galaxy.far.far.away}

\newcommand{\simgt}{\,\rlap{\lower 3.5 pt \hbox{$\mathchar \sim$}} \raise 1pt \hbox {$>$}\,}
\newcommand{\simlt}{\,\rlap{\lower 3.5 pt \hbox{$\mathchar \sim$}} \raise 1pt \hbox {$<$}\,}
\newcommand{\dd}{\mathrm{d}}

\newcommand{\BE}{\begin{equation}}
\newcommand{\EE}{\end{equation}}
\newcommand{\BEA}{\begin{eqnarray}}
\newcommand{\EEA}{\end{eqnarray}}

\newcommand{\Ob}{\Omega_\textrm{b}}
\newcommand{\Om}{\Omega_\textrm{m}}
\newcommand{\OL}{\Omega_\Lambda}
\newcommand{\rhoc}{\rho_\textrm{c}}

\newcommand{\CHII}{C_\textrm{\ion{H}{2}}}
\newcommand{\DV}{\ifmmode{\Delta v_{\mathrm{Ly}\alpha}}\else $\Delta v_{\mathrm{Ly}\alpha}$\xspace\fi}

\newcommand{\Tigm}{{\mathcal{T}_\textsc{igm}}}

\newcommand{\HI}{\ifmmode{\textsc{hi}}\else H\textsc{i}\fi\xspace}
\newcommand{\HII}{\ifmmode{\textsc{hii}}\else H\textsc{ii}\fi\xspace}
\newcommand{\OII}{[O\textsc{ii}]}
\newcommand{\OIII}{O\textsc{iii}}
\newcommand{\CIV}{C\textsc{iv}}
\newcommand{\HeII}{He\textsc{ii}}

\newcommand{\Msun}{\ifmmode{M_\odot}\else $M_\odot$\xspace\fi}
\newcommand{\MUV}{\ifmmode{M_\textsc{uv}}\else $M_\textsc{uv}$\xspace\fi}
\newcommand{\fesc}{\ifmmode{f_\textrm{esc}}\else $f_\textrm{esc}$\xspace\fi}
\newcommand{\lya}{\ifmmode{\mathrm{Ly}\alpha}\else Ly$\alpha$\xspace\fi}
\newcommand{\Ha}{\ifmmode{H$\alpha$}\else $H\alpha$\xspace\fi}
\newcommand{\vrot}{v_\textrm{rot}}

\newcommand{\nh}[1][]{\ifmmode{\overline{n}_\textsc{h}^{#1}}\else $\overline{n}_\textsc{h}$\xspace\fi}

\newcommand{\xHI}{\ifmmode{x_\HI}\else $x_\HI$\xspace\fi}
\newcommand{\xHImean}{\ifmmode{\overline{x}_\HI}\else $\overline{x}_\HI$\xspace\fi}
\newcommand{\xHIImean}{\ifmmode{\overline{x}_\HII}\else $\overline{x}_\HII$\xspace\fi}
\newcommand{\trec}{\ifmmode{t_\textrm{rec}}\else $t_\textrm{rec}$\xspace\fi}
\newcommand{\clump}[1][]{\ifmmode{C_\HII^{#1}}\else $C_\HII$\xspace\fi}
\newcommand{\xiion}{\ifmmode{\xi_\mathrm{ion}}\else $\xi_\mathrm{ion}$\xspace\fi}

\newcommand{\Nion}{\ifmmode{\dot{N}_{\mathrm{ion}}}\else $\dot{N}_\mathrm{ion}$\xspace\fi}
\newcommand{\Rion}[1][]{\ifmmode{R_\mathrm{ion}^{#1}} \else $R_\mathrm{ion}$\xspace\fi}

\newcommand{\Tb}{\ifmmode{T_{21}}\else $T_{21}$\xspace\fi}
\newcommand{\aesc}{\ifmmode{\alpha_\mathrm{esc}}\else $\alpha_\mathrm{esc}$\xspace\fi}
\newcommand{\fescII}{\ifmmode{f_\mathrm{esc,10}^\textsc{ii}}\else $f_\mathrm{esc,10}^\textsc{ii}$\xspace\fi}
\newcommand{\fescIII}{\ifmmode{f_\mathrm{esc,7}^\textsc{ii}}\else $f_\mathrm{esc,7}^\textsc{iii}$\xspace\fi}
\newcommand{\astarII}{\ifmmode{\alpha_\star^\textsc{ii}}\else $\alpha_\star^\textsc{ii}$\xspace\fi}
\newcommand{\astarIII}{\ifmmode{\alpha_\star^\textsc{iii}}\else $\alpha_\star^\textsc{iii}$\xspace\fi}
\newcommand{\fstarII}{\ifmmode{f_{\star,10}^\textsc{ii}}\else $f_{\star,10}^\textsc{ii}$\xspace\fi}
\newcommand{\fstarIII}{\ifmmode{f_{\star,7}^\textsc{iii}}\else $f_{\star,7}^\textsc{iii}$\xspace\fi}
\newcommand{\tstar}{\ifmmode{t_\star}\else $t_\star$\xspace\fi}
\newcommand{\Mturn}{\ifmmode{M_\mathrm{turn}}\else $M_\mathrm{turn}$\xspace\fi}
\newcommand{\LX}{\ifmmode{L_X/{\dot{M}_\star}}\else $L_X/{\dot{M}_\star}$\xspace\fi}
\newcommand{\nuX}{\ifmmode{E_0}\else $E_0$\xspace\fi}
\newcommand{\AVCB}{\ifmmode{A_\mathrm{VCB}}\else $A_\mathrm{VCB}$\xspace\fi}
\newcommand{\ALW}{\ifmmode{A_\mathrm{LW}}\else $A_\mathrm{LW}$\xspace\fi}
\newcommand{\Mpcinv}{\ifmmode{\,\mathrm{Mpc}^{-1}}\else \,Mpc$^{-1}$\xspace\fi} 

\newcommand{\kp}{\ifmmode{k_\textrm{peak}}\else $k_\textrm{peak}$\xspace\fi}
\newcommand{\hp}{\ifmmode{h_\textrm{peak}}\else $h_\textrm{peak}$\xspace\fi}
\newcommand{\hMpc}{\ifmmode{\,h^{-1}\textrm{Mpc}}\else \,$h^{-1}$Mpc\xspace\fi}

\newcommand{\fdens}{\,erg s$^{-1}$ cm$^{-2}$\xspace}
\newcommand{\kms}{\,\ifmmode{\mathrm{km}\,\mathrm{s}^{-1}}\else km\,s${}^{-1}$\fi\xspace}
\newcommand{\cm}{\,\ifmmode{\mathrm{cm}}\else cm\fi\xspace}

\newcommand{\HST}{\textit{HST}}
\newcommand{\JWST}{\textit{JWST}}
\newcommand{\WFIRST}{\textit{WFIRST}}

\newcommand{\NB}[1]{\textbf{\color{red} #1}}
\newcommand{\tnm}[1]{$^\textrm{#1}$}

\title{CANUCS: An Updated Mass and Magnification Model of Abell 370 with JWST}

\correspondingauthor{Rachel Gledhill}
\email{rlq450@alumni.ku.dk}


\author{Rachel Gledhill}
\affiliation{Cosmic Dawn Center (DAWN), Denmark}
\affiliation{Niels Bohr Institute, University of Copenhagen, Jagtvej 128, DK-2200 Copenhagen N, Denmark}

\author[0000-0002-6338-7295]{Victoria Strait}
\affiliation{Cosmic Dawn Center (DAWN), Denmark}
\affiliation{Niels Bohr Institute, University of Copenhagen, Jagtvej 128, DK-2200 Copenhagen N, Denmark}

\author[0000-0001-8325-1742]{Guillaume Desprez}
\affiliation{Department of Astronomy and Physics and Institute for Computational Astrophysics, Saint Mary's University, 923 Robie Street, Halifax, Nova Scotia B3H 3C3, Canada}

\author[0009-0009-4388-898X]{Gregor Rihtar{\v s}i{\v c}}
\affiliation{University of Ljubljana, Department of Mathematics and
Physics, Jadranska ulica 19, SI-1000 Ljubljana, Slovenia}

\author[0000-0001-5984-0395]{Maru{\v s}a Brada{\v c}}
\affiliation{University of Ljubljana, Department of Mathematics and
Physics, Jadranska ulica 19, SI-1000 Ljubljana, Slovenia}
\affiliation{Department of Physics and Astronomy, University of California Davis, 1 Shields Avenue, Davis, CA 95616, USA}

\author[0000-0003-2680-005X]{Gabriel Brammer}
\affiliation{Cosmic Dawn Center (DAWN), Denmark}
\affiliation{Niels Bohr Institute, University of Copenhagen, Jagtvej 128, DK-2200 Copenhagen N, Denmark}

\author[0000-0002-4201-7367]{Chris J. Willott}
\affiliation{National Research Council of Canada, Herzberg Astronomy \& Astrophysics Research Centre, 5071 West Saanich Road, Victoria, BC, V9E 2E7, Canada}

\author[0000-0003-3243-9969]{Nicholas Martis}
\affiliation{University of Ljubljana, Department of Mathematics and
Physics, Jadranska ulica 19, SI-1000 Ljubljana, Slovenia}

\author[0000-0002-7712-7857]{Marcin Sawicki}
\affiliation{Department of Astronomy and Physics and Institute for Computational Astrophysics, Saint Mary's University, 923 Robie Street, Halifax, Nova Scotia B3H 3C3, Canada}

\author{Gaël Noirot}
\affiliation{Department of Astronomy and Physics and Institute for Computational Astrophysics, Saint Mary's University, 923 Robie Street, Halifax, Nova Scotia B3H 3C3, Canada}

\author[0000-0001-8830-2166]{Ghassan T. E. Sarrouh}
\affiliation{Department of Physics and Astronomy, York University, 4700 Keele St., Toronto, Ontario, Canada, MJ3 1P3}

\author[0000-0002-9330-9108]{Adam Muzzin}
\affiliation{Department of Physics and Astronomy, York University, 4700 Keele St., Toronto, Ontario, Canada, MJ3 1P3}

\begin{abstract}
We report an updated mass and magnification model of galaxy cluster Abell 370 using new NIRCam and NIRISS data from the CAnadian NIRISS Unbiased Cluster Survey (CANUCS). 
Using \texttt{Lenstool} and a combination of archival \textit{HST} and MUSE data with new \textit{JWST} data as constraints, we derive an improved gravitational lensing model and extract magnifications of background galaxies with uncertainties. Using our best fit model, we perform a search for new multiply imaged systems via predicted positions. 
We report no new multiply imaged systems with identifiable redshifts, likely due to already very deep \textit{HST} and \textit{Spitzer} data, but confirm a $z\sim8$ multiply imaged system by measuring its redshift with NIRISS and NIRSpec spectra.
We find that the overall shape of the critical curve for a source at $z = 9.0$ is similar to previous models of Abell 370, with small changes. We investigate the $z\sim8$ galaxy with two images observable with an apparent magnitude in the F125W band of $26.0\pm0.2$ and $25.6\pm0.1$. After correcting for the magnifications of the images, 7.2$^{+0.2}_{-1.2}$ and 8.7$^{+0.4}_{-0.4}$, we use SED fitting to find an intrinsic stellar mass of log($M^*/M_{\odot})$ = 7.35$^{+0.04}_{-0.05}$, intrinsic SFR of 3.5$^{+2.2}_{-1.4}$ M$_{\odot}$/yr, and $M_{UV}$ of -21.3$^{+0.2}_{-0.2}$, which is close to the knee of the luminosity function at that redshift. Our model, and corresponding magnification, shear, and convergence maps are available on request and will be made publicly available on MAST in a CANUCS data release (DOI: 10.17909/ph4n-6n76).

\end{abstract}


\section{Introduction} \label{sec:intro}

Mass and magnification models of galaxy clusters have now been a mainstay in astronomy for many years, because of their utility in determining the magnification of high redshift galaxies that would otherwise be too faint to detect or get deep spectroscopy (e.g., \citealp{Richard2010, Hashimoto2018,Strait2018,Welch2022, Bergamini2023,Vanzella2023,Adamo2024arXiv,Bradac2024,Fujimoto2024arXiv,Mowla2024arXiv}). This is of particular use for studying fainter galaxies at fixed redshift, which are more representative of characteristic galaxies at that epoch than the more luminous contemporary galaxies detected without the benefit of gravitational lensing. They are important to study for a full understanding of galaxy growth and evolution (e.g., \citealp{Williams2022, Asada2022, Strait2023, Atek2023}). 

The most massive clusters generally have the largest areas of high magnification, a desirable quality when investigating background galaxies \citep{Lotz2017}. Additionally, the best-constrained models are generally those that contain the highest number of spectroscopically confirmed multiply imaged systems (e.g. \citealp{Bergamini2023}), so confirming redshifts of multiply imaged background sources is a high priority. 

Abell 370 has the longest history of strong lensing discoveries of any galaxy cluster, with the first detection of a strongly lensed object behind Abell 370 at $z=0.724$ by \textcite{Soucail1987}. Because of the cluster's large mass ($\sim10^{14} M_{\odot}$) and high redshift ($z=0.375$), it was selected as a part of the Hubble Frontier Fields Survey \citep{Lotz2017} and has had many mass and magnification modelling efforts performed since, ranging in method and available data \citep{hammer1987, Richard2010, Medezinski2011, Umetsu2011, johnson2014, Diego2018, Lagattuta2017, kawamata2017, Strait2018, Lagattuta2019, Lagattuta2022}.

In this paper, we present an updated parametric lens model from \texttt{Lenstool} (\citealp{Kneib1993}, \citealp{Jullo2007}, \citealp{Jullo2009}) using new \textit{JWST} NIRCam, NIRSpec, and NIRISS data from the CAnadian NIRISS Unbiased Cluster Survey (CANUCS) \citep{Willott2022} in addition to the wealth of archival data from \textit{Hubble Space Telescope }\textit{(HST)} and the Multi-Unit Spectroscopic Explorer (MUSE). We report the mass and magnification distribution products for public use in the same format as the Hubble Frontier Field (HFF) clusters. We also report the spectroscopic confirmation of a multiply imaged $z\sim8$ galaxy with \textit{JWST}/NIRSpec. 

The structure of the paper is as follows: In Section \ref{sec:data} we explain the data we used in the lens model. In Section \ref{sec:methods} we describe the process used for the lens model and in Section \ref{sec:res} we report our magnification distribution and some properties of the $z=8$ multiple images derived through stellar population fitting. We discuss these and finally conclude in Section \ref{sec:disc}. We assume a $\Lambda$CDM cosmology with $H_{0}$ = 70 km s$^{-1}$ Mpc$^{-1}$, $\Omega_{m}$ = 0.3, and
$\Omega_{\Lambda}$ = 0.7. All magnitudes are AB magnitudes.

\section{Data} \label{sec:data}
We utilize data from the Abell 370 cluster field obtained via CANUCS \citep{Willott2022}, a \textit{JWST} NIRISS GTO program. We use \textit{JWST}/NIRISS and \textit{JWST}/NIRSpec data and all available NIRCam and \textit{HST} imaging for identification and spectroscopic confirmation of some multiply imaged systems. We also make use of significant archival data from MUSE \citep{Lagattuta2017, Lagattuta2019, Lagattuta2022} and \textit{HST} \citep{Lotz2017}. 

\subsection{Imaging and Photometry}\label{sec:imaging}
In this paper, we used imaging data to identify multiply imaged sources. The Abell 370 cluster field was observed with \textit{JWST}/NIRCam filters F090W, F115W, F150W, F200W, F277W, F356W, F410M, and F444W with exposure times of 6.4 ks each, reaching a signal to noise ratio of $\sim5-10$ for a point source at AB magnitude of 29. We also use archival \textit{HST} imaging from the Hubble Frontier Fields program as well as others (PIDs 14216, 14038, 15117, 13459, 13790, 11108, 11591, 14209, 11507, 15940, 11582) in F105W, F125W, F140W, F160W, F435W, F606W, and F814W as a part of the multiple image identification process, as well as spectral energy distribution (SED) fitting for the $z\sim8$ source discussed in Section \ref{sec:z8mis}. 

All \textit{HST} and \textit{JWST} imaging was reduced and analyzed following a similar process as is described by \cite{Noirot2023}, which accounts for background subtraction, removal of brightest cluster galaxies (BCGs) from the cluster field \citep{Martis2024}, point spread function (PSF) matching, and techniques to remove the various sources of noise in NIRCam data detailed by \cite{Rigby2023}. In this paper, we assume a 0$\farcs 5$ aperture for photometry of background galaxies.

\subsection{NIRISS Spectroscopy}\label{sec:niriss}
NIRISS spectroscopy was used for two systems that had previously uncertain redshifts (Systems 9 and 11). The NIRISS spectroscopy in Abell 370 consists of wide-field slitless spectroscopy in two orientations for each of F115W, F150W, and F200W (9.6 ks for each filter/orientation combination). The reduction and contamination modelling for the NIRISS spectroscopy is performed with \texttt{grizli}\footnote{github.com/gbrammer/grizli} (DOI: 10.5281/zenodo.1146904), also following closely the process of \cite{Noirot2023}.

\subsection{NIRSpec Spectroscopy}\label{sec:spec}
We used NIRSpec multi-object prism spectroscopy for three systems that previously had only photometric redshifts available, Systems 8, 11, and 43. The NIRSpec spectra were processed in the manner described in \textcite{Desprez2023}. When images from other systems are captured in both the available NIRSpec data and the MUSE data for this cluster, the redshifts are consistent with each other.  

\subsection{Archival MUSE Data}\label{sec:archival}
Finally, we take advantage of the wealth of MUSE data first reported by \cite{Lagattuta2017} and subsequently by \cite{Lagattuta2019, Lagattuta2022}. There is a combination of deep (2-8 hours) and shallow (1 hour) MUSE pointings, which cover the entire cluster field. As an integral field unit, MUSE blindly recovers emission and provides the bulk of the spectroscopic redshifts in Abell 370, including for 8 systems which we cannot see in \textit{HST} or \textit{JWST} (which we exclude in our main model). The catalogs we used for the analysis are from the supplementary materials of \cite{Lagattuta2019}\footnote{https://academic.oup.com/mnras/article/485/3/3738/ 5368372\#supplementary-data}.

\section{Methods}  \label{sec:methods}

\subsection{Multiple Image System Identification}\label{MI-id}

\begin{figure*}
    \centering
    \includegraphics[width=0.75\textwidth]{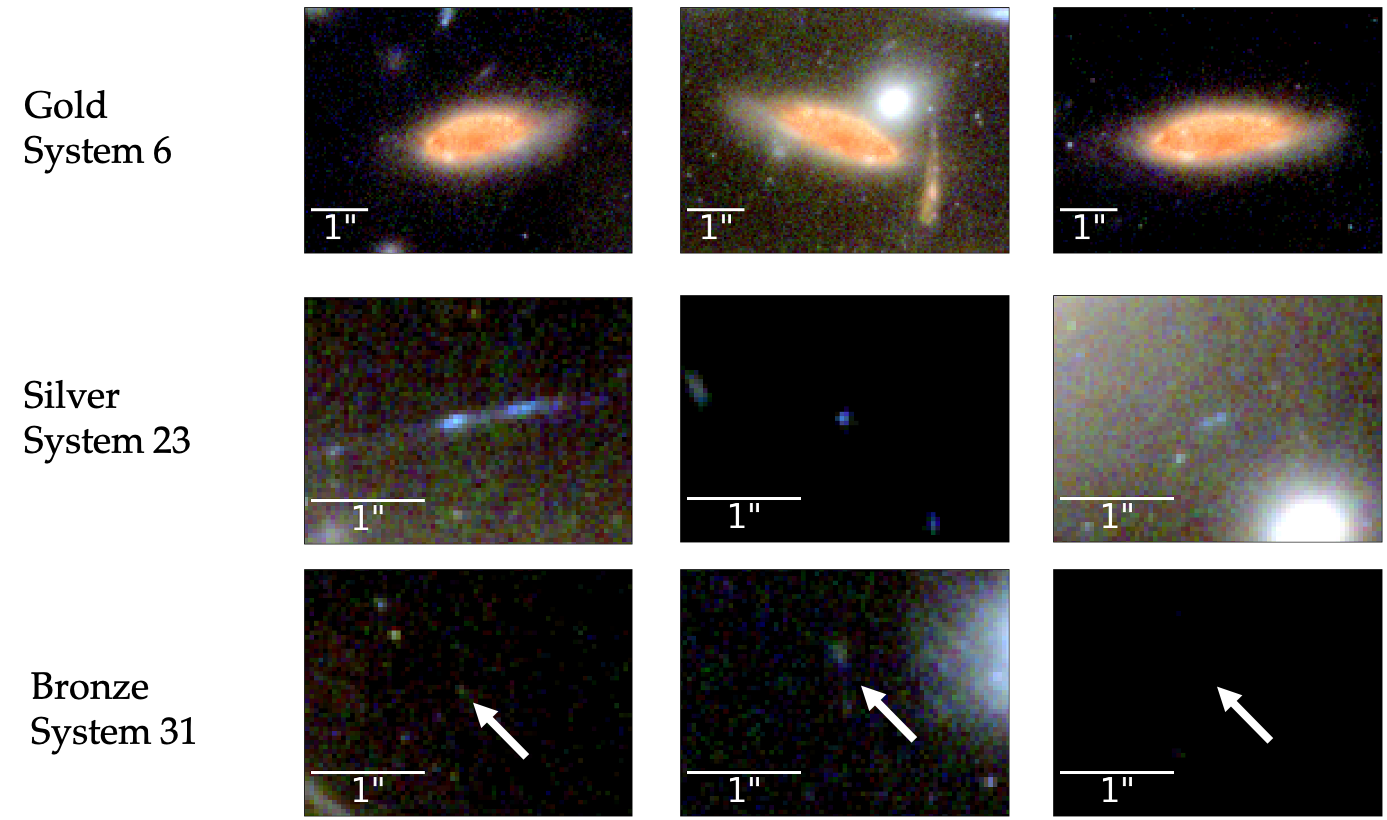}
    \caption{Examples of Gold, Silver, and Bronze systems shown in an RGB image made of F200W, F150W, and F115W filters. \textbf{Top}: System 6 is a Gold system - it has 3 images with well-determined redshifts and the colors and morphology of the images make it clear that they are the same galaxy. \textbf{Middle}: System 23 is a Silver system; there are 2 knots visible in some images but not the others and the colors of the images appear slightly different. The Ly-$\alpha$ emission that determines its redshift is much stronger in the first image than the other two. This may be a true system, but because of these factors, we are not confident enough to include it in the Gold category. \textbf{Bottom}: System 32 is a Bronze system; while these images all show Ly-$\alpha$ emission at the same redshift in the MUSE data, the images are very faint and the colours appear somewhat different. The third image (right) is an example of a Quartz image - there is no source seen here in the NIRCAM data. This system is included in the catalogue for completeness but is classified as Bronze because we cannot confidently verify that it is a true system. }
    \label{fig:goldsilverbronze}
\end{figure*}

\begin{figure*}
    \centering
    \includegraphics[width=0.7\textwidth]{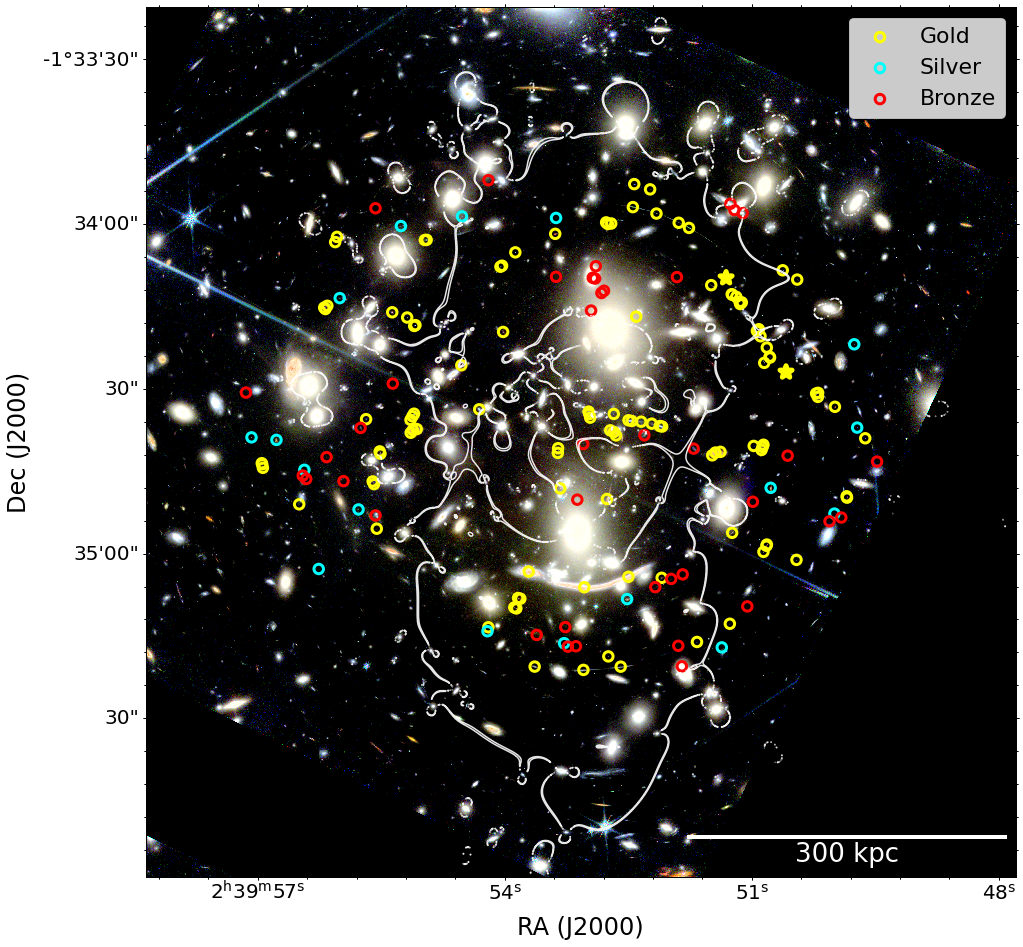}
    \caption{A F200W-F150W-F115W RGB image of the cluster with a critical curve at $z=9.0$. The multiple images systems are plotted and color-coded by whether they are a member of the gold (yellow circles), silver (cyan circles), and bronze (red circles) catalogues. The two System 11 images are marked with yellow stars. }
    \label{fig:cc}
\end{figure*}

We use the \textit{JWST} data to re-examine multiply imaged systems identified in previous works (\citealp{Strait2018}, \citealp{Lagattuta2019}) and construct a robust sample of spectroscopically confirmed multiple image systems. The primary criteria for a high-confidence system are that its redshift is spectroscopically confirmed for all its images and that the colours of each source image in the NIRCAM data match the other images in all filters. We made 3 catalogs, which we labelled Gold, Silver, and Bronze. Gold contains systems that fulfill the criteria; the morphology and colour of the individual images match each other and we have high confidence in the spectroscopic redshift of the images. A Silver system is likely a true multiply imaged system but may be lacking in confidence in the redshift or have differences in morphology and colour between the images. A Bronze system is one that is included because it is a known possible system, but there is sufficient doubt in its redshift or in whether the images belong to the same source that it was not considered for inclusion in this model. Examples are shown in Fig. \ref{fig:goldsilverbronze} and the positions of each system are plotted in Fig. \ref{fig:cc}. Some systems show clear MUSE detections but show no visible counterpart in HST or JWST images. Following the example of \textcite{Niemiec2023}, we classified these systems as `Quartz' as they are disqualified from the Gold catalogue because we are unable to fully evaluate them rather than being found insufficiently robust on examination. The main model was constructed using only Gold systems.  

When the redshift determination from the MUSE data was inconclusive, we extracted spectra for these sources from the NIRISS data. For System 9, H$\alpha$ and [OIII]5007+4960 emission lines in the NIRISS spectra give a spectroscopic redshift of $z=1.514\pm0.004$. This provides an independent spectroscopic redshift measurement which is consistent with the higher-resolution spec-z from \textcite{Lagattuta2019}, $z=1.5182\pm0.0001$, allowing us to upgrade this system to the Gold catalogue given our criteria. 

We looked at a z$\sim$8 system (System 11) previously discovered by \textcite{Strait2018}. Prior to the availability of NIRSpec data, we were able to confirm its redshift with NIRISS. Processing its NIRISS spectrum with \texttt{grizli}, we are able to identify the Lyman break in the spectra of both images of the galaxy, giving a grism redshift of $z=7.59\pm0.04$. This system's redshift was later determined more precisely with NIRSpec to be $z=7.6476\pm0.0006$. Another system that previously had only a photometric redshift is System 8. Its NIRSpec spectrum revealed it has a spectroscopic redshift of $z=2.866\pm0.003$, quite close to the photometric redshift, $z=2.98$ \citep{Strait2018}. 

\begin{figure}
    \centering
    \includegraphics[width=0.47\textwidth]{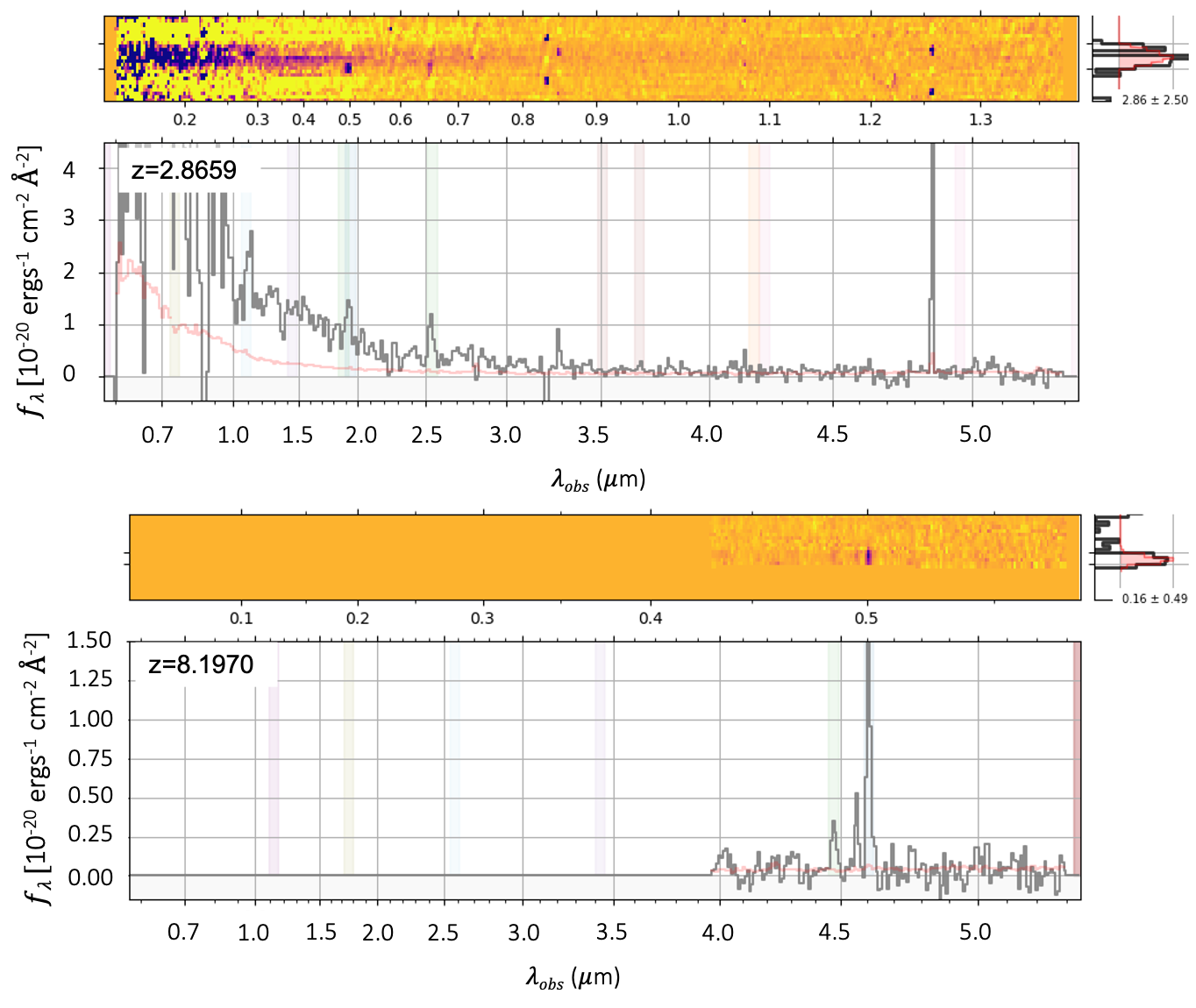}
    \caption{NIRSpec spectra of Image 8.2 (top) and Image 43.2 (bottom). }
    \label{fig:sys8_43}
\end{figure}

A third system in the multiple image catalogue for which we were able to update the redshift with NIRSpec is System 43. The colours and number of knots of its two images match in both JWST and HST data but it previously had no redshift measurement. Image 43.2 has a NIRSpec redshift of $z=8.1970\pm0.0007$, which is consistent with \textcite{Lagattuta2019}'s estimate based on their lensing model. We did not include it as a constraint in our model because only one image has a spectroscopic redshift. The spectra of System 8 and 43 are shown in Fig. \ref{fig:sys8_43}.

When individual features or `knots' in a galaxy were visible in all images of a system in the NIRCam data, these were separated into subsystems when lens modelling. In some systems, features in a single stretched image of a galaxy are separated by more than 5\arcsec. 


The full catalogue of multiple images is found in Appendix \ref{app:mi}.

\subsection{Cluster Member Selection}\label{sec:cm-select}

\begin{figure}
    \centering
    \includegraphics[width = 0.85\linewidth]{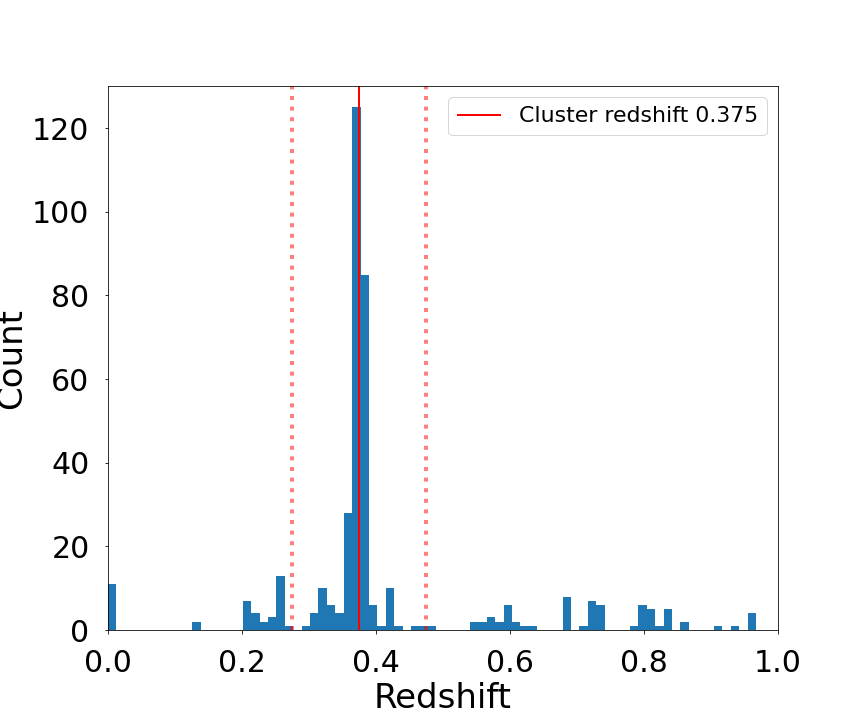}
    \caption{Histogram of MUSE redshifts for objects in the field. The solid line is the cluster redshift 0.375 and the dotted lines show the $\pm$0.1 range from which cluster members were selected.}
    \label{fig:cmhist}
\end{figure}

\begin{figure*}
    \centering
    \includegraphics[width = 0.73\textwidth]{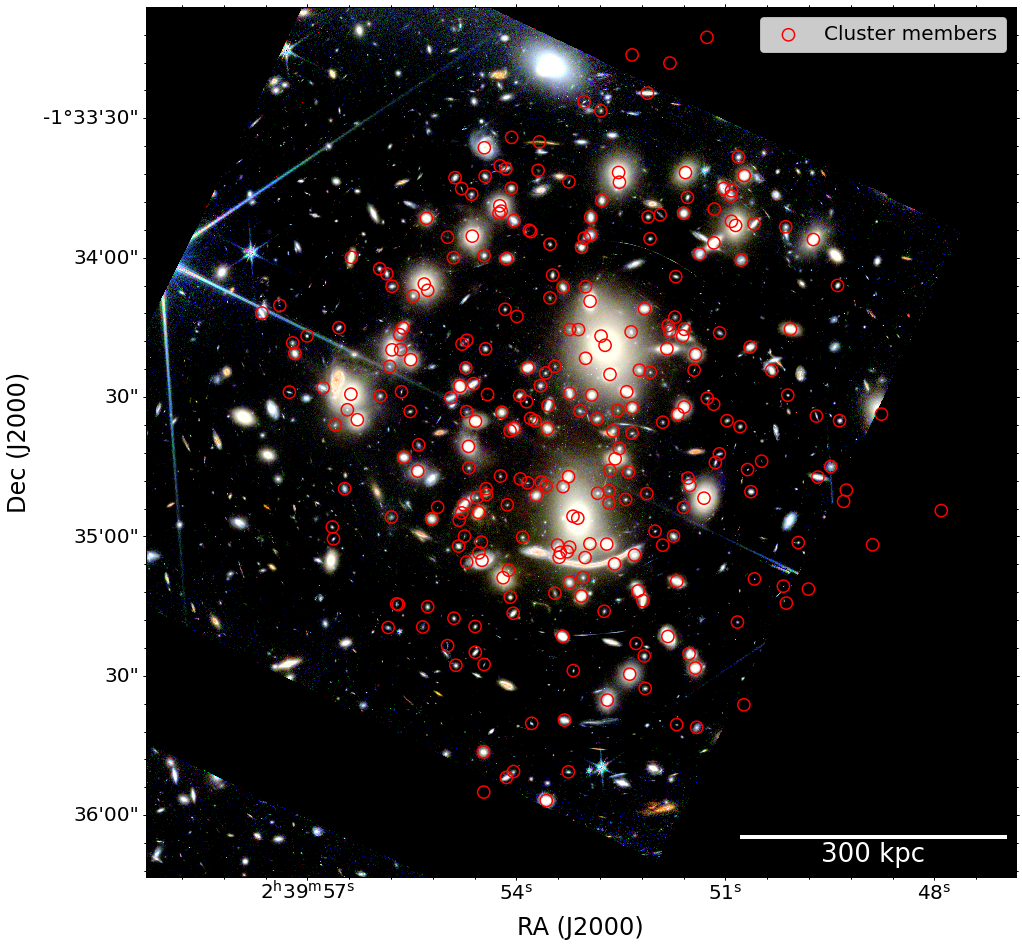}
    \caption{NIRCAM F200W-F150W-F115W RGB image of Abell 370 with the 280 spectroscopically confirmed cluster member galaxies plotted as red circles.}
    \label{fig:clustermembers}
\end{figure*}

We select the cluster member galaxies using the catalog of objects with redshifts derived from MUSE observations from \textcite{Lagattuta2019}. Galaxies within a range of $\pm$0.1 of the cluster redshift 0.375 are taken as cluster members, for a total of 280 galaxies; the redshift distribution with a peak around the cluster redshift is shown in Fig. \ref{fig:cmhist}. The positions of the cluster member galaxies are plotted in Fig. \ref{fig:clustermembers}. The brightest cluster member has an apparent magnitude in the F814W band of 19.8 and the faintest has an apparent magnitude of 30.7. 

\subsection{Lenstool}\label{sec:lenstool}

We used the publicly available parametric lensing code \texttt{Lenstool} to model the mass and magnification distribution of the cluster. In \texttt{Lenstool}, the mass in Abell 370 is modelled by individual components with analytic mass density profiles. We use a pseudo-isothermal elliptic mass distribution (PIEMD) profile \citep{kassiola1993} for each component. \texttt{Lenstool} uses the positions of the multiple image systems as constraints and minimizes the distance between the positions predicted by the model and the true image positions using a Monte Carlo Markov Chain sampler. Here, we used the Gold catalogue of image systems - 24 systems containing a total of 63 images. 12 of the systems contain knots. \texttt{Lenstool} takes coordinates, ellipticity, orientation angle, core radius, cut radius, and central velocity dispersion as parameters. Cluster-scale halos are placed in the cluster to model the larger distribution of dark matter and intracluster gas, while galaxy-scale halos are placed at the position of the cluster member galaxies. The cut radius is fixed to a large value of 800 kpc for the cluster-scale halos because for mass halos on the scale of the cluster's size, this parameter is expected to be larger than the strong lensing radius of the cluster and therefore cannot be constrained with our data (\citealp{Lagattuta2019}).  

The galaxy-scale halos are fixed to the positions of the cluster-member galaxies. The ellipticity parameter is eliminated by modelling the cluster members as circular for ease of calculation. A test using a subset of the catalogue for which ellipticity and angle measurements were available \citep{Strait2018} showed only a 4\% difference in $\chi^2$ when fit with circular cluster members compared to their true ellipticities. The core radius, cut radius, and velocity dispersion of these halos are scaled according to their luminosities using the relations described in \textcite{Kneib1996}, which correspond to a constant mass-to-light ratio. The scaling parameters can be either fixed or be fit by \texttt{Lenstool}. For this work, the scaling is core radius is fixed at 0.15 kpc, while the scaling cut radius is fit by \texttt{Lenstool} with a uniform prior between 10 and 50 kpc. The scaling value for the velocity dispersion has a Gaussian prior with a mean of 158 km/s and a standard deviation of 27 km/s \citep{Lagattuta2019}. The reference magnitude when scaling all parameters is 20.701. 

Three cluster member galaxies were not modelled with the scaling relations but as separate PIEMD halos. These were the 2 BCGs and a smaller galaxy that had a large influence on the fit of System 2 due to its proximity to the system. The ellipticity and angle parameters are fixed based on the cluster members catalogue from \textcite{Strait2018} and the core radius remains fixed at 0.15 kpc like the rest of the cluster members. The cut radius and velocity dispersion are allowed to vary. 

We started with a simple initial guess of 2 cluster-scale halos centered on the brightest cluster galaxies (BCGs) to account for the dark matter in the cluster. We added additional halos with wide optimization limits one at a time and iterated adjusting and narrowing the limits until the $\chi^2$ statistic stopped improving. We also examined the reconstructions of individual systems for each configuration of halos and when the fit was consistently worse than average in a region of the cluster, a halo was added to that region. The best-fit model is described in Section \ref{sec:model}.  

\begin{figure*}
    \centering
    \includegraphics[width=0.82\textwidth]{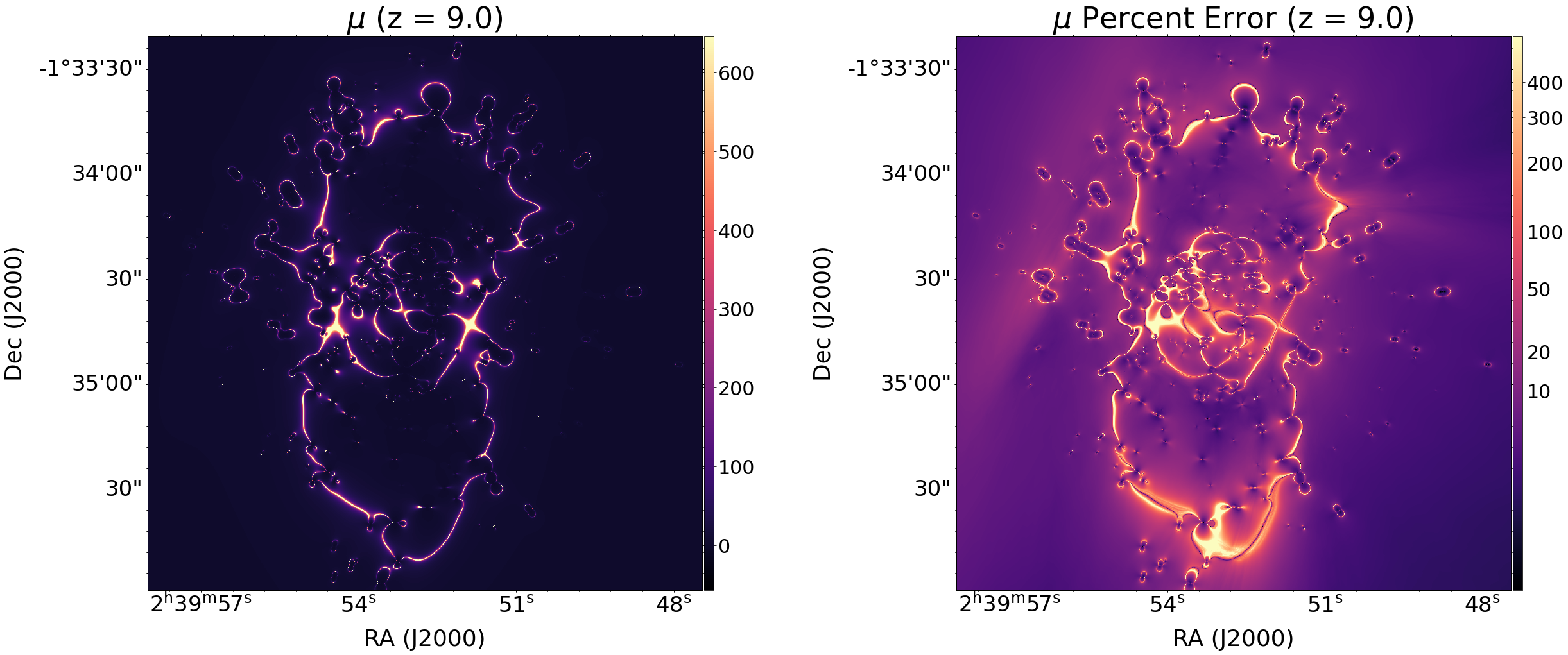}
    \caption{Left: The magnification map for a source at $z = 9.0$. Right: The relative uncertainty $\Delta \mu/\mu$ on the magnification of a source at $z = 9.0$, expressed as a percentage. Note that the uncertainty becomes very large for the most magnified objects near the critical curve, but is lower elsewhere. It is not advisable to use the magnification estimates without considering whether the error is reasonable at those coordinates.}
    \label{fig:mag_err}
\end{figure*}

We derived uncertainties in magnification by generating magnification maps for the last 100 iterations performed by \texttt{Lenstool} during its run. The errors on magnification were taken as the difference between the 16th and 84th percentile of magnification values in each pixel. The magnification map and corresponding magnification error percentage map for a source at $z=9.0$ are shown in Fig. \ref{fig:mag_err}.  

Finally, we used our model to try to identify new multiple-image system candidates. To this end, we ran \texttt{Lenstool} with our current best model and predicted potential counter images to all sources in the cluster field, brighter than the 27th magnitude in the F200W band, using their photometric redshift. We then visually inspected the region around each predicted counter image and identified potential candidates based on color similarity in available bands and photometric redshift similarity. We did not find any potential new candidates reliable enough to include in our catalogues.

\section{Results} \label{sec:res}
\subsection{Lens Model}\label{sec:model}

The final best-fit lens model of Abell 370 was constrained by the Gold multiple image systems and features 5 cluster-scale halos and 280 galaxy-scale halos. It was found by starting with the galaxy-scale halos and two cluster-scale halos centered on the BCGs, providing a wide parameter space for optimization and iterating by adjusting and narrowing the limits of the parameter space until the model stabilized. We tested adding additional halos in regions where the reconstructions of image positions were poor, providing wide limits and iterating again as long as the model was improving in $\chi^2$. 

\begin{figure*}
    \centering
    \includegraphics[width=0.8\textwidth]{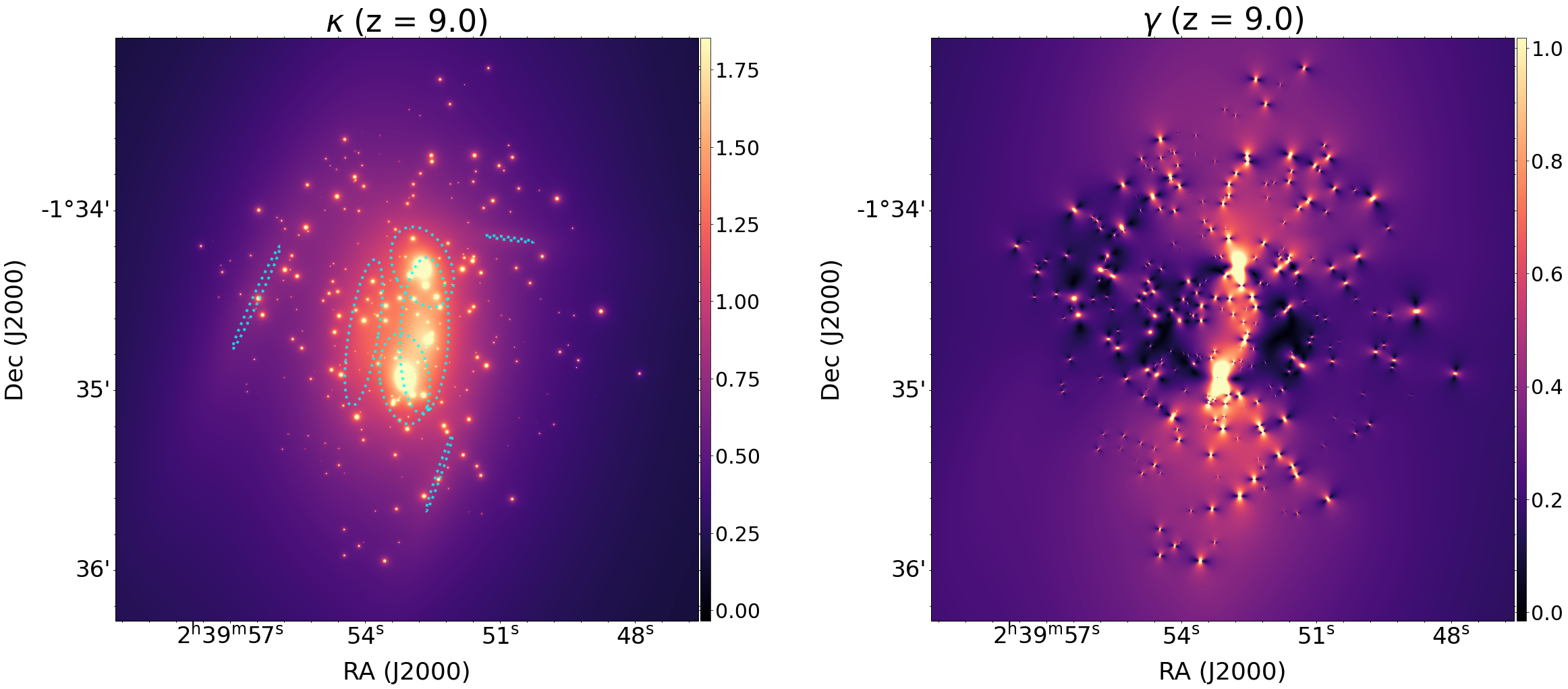}
    \caption{Left: The convergence $\kappa$ found by our model for a source at $z = 9.0$. Ellipses show the location of the cluster-scale halos. Note that the sizes of these ellipses are not physical sizes. Instead, the size of each ellipse is scaled to show the halo's velocity dispersion (a proxy for its mass) relative to the other halos. Right: The shear $\gamma$ for a source at $z = 9.0$.}
    \label{fig:kappagamma}
\end{figure*}

Two halos are located near the centre of the cluster, bridging the distance between the two BCGs. Another halo on the eastern side of the cluster near the arc of cluster member galaxies sometimes referred to as the `crown' and a halo to the south of the cluster centre are located near over-concentrations in cluster member galaxies. A third small halo in the northeast region is found in an under-density in visible mass of the cluster. While the exact number of halos and their placement differ, there are some similarities to the \textcite{Lagattuta2019} model, particularly in the placement of the crown halo and in one of the central halos being shifted to the east relative to the two BCGs. The critical curve of a source at $z=9.0$ is shown in Fig. \ref{fig:cc}. The convergence ($\kappa$) and shear ($\gamma$) for a source at $z = 9.0$ are displayed in Fig. \ref{fig:kappagamma}. 

The model was evaluated using the $\chi^2$ statistic output by \texttt{Lenstool}. The rms distance between the reconstructed and true image positions of each system was also calculated to compare the ability of the model to describe each system. The statistics for the best-fit model are described in Table \ref{tab:fit_stats}. 

\begin{table}[h]
\centering
\caption{Statistics describing the fit of the best lens-model. The $\chi^2$ and evidence are provided as an output by \texttt{Lenstool}, while the delta-rms deviation was calculated as the rms deviation of all the separations between the images reconstructed by \texttt{Lenstool} and the actual positions of the images.} 
\begin{tabular}{ l | l }
 Statistic & Value \\ 
 \hline \hline
 Reduced chi-squared $\chi^2/\nu$ & 3.03 \\ 
 Degrees of freedom $\nu$ & 102 \\
 Bayesian information criterion & 496.92\\
 Delta-rms & $0\farcs 48$ \\
 &
\end{tabular}
\label{tab:fit_stats}
\end{table}

\subsection{Problem systems}\label{sec:problem}

While the majority of the system images are well predicted by the model, there are a few systems where the fit is poor. 

For some systems, such as Systems 3 and 5, compact images were well predicted by the model, while \texttt{Lenstool} struggled to reconstruct the more extended arcs of the system, where reconstructed images may be up to 7 arcseconds away from the true image. 

System 7 is a problem system for this and previous models (\citealp{Strait2018}, \citealp{Lagattuta2019}). It has up to 7 images depending on the Abell 370 multiple image catalogue we consult and is sometimes also separated into 2 different systems. Note that this is the reason there is no System 10 in our catalogue. To keep the system numbers consistent with other works (\citealp{Strait2018}, \citealp{Lagattuta2019}), its images have been folded into System 7 in Appendix \ref{app:mi}. For this work, we have included images 7.1 and 7.6, which are sometimes separated into 2 images each as single images in our Gold catalogue, and included the other possible images in the Bronze catalogue on the grounds that their redshifts in the MUSE data are not as well determined.

\subsection{Stellar Population Fitting for z=7.6 Multiple Images}\label{sec:z8mis}
\label{sec:system11}

We used Bayesian Analysis of Galaxies for Physical Inference and Parameter EStimation (BAGPIPES) to model the stellar properties of System 11, a multiply imaged $z\sim8$ galaxy. BAGPIPES is a Python-based SED fitting code designed to model the continuum and nebular emission of galaxies \citep{Carnall2018}. It generates model galaxy spectra based on the BC03-MILES stellar population synthesis models (\citealp{Bruzal03}, \citealp{Falcon-Barroso2011}), varying age and metallicity in a grid. BAGPIPES uses the initial mass function defined by \textcite{Kroupa2002}. We use the \textcite{Calzetti2000} dust law, which is often used for high-redshift galaxies as it is based on dust attenuation in local galaxies with active star formation (e.g. \citealp{Carnall2018}, \citealp{Strait2021}). We also ran BAGPIPES assuming constant star formation.

The priors we used when fitting with BAGPIPES are described in Table \ref{tab:allpriors}. 

\begin{table}[h]
\centering
\caption{Priors of the model provided to \textsc{bagpipes} used when fitting the spectrum of the System 11 images. Parameters labelled Const. are specific to the constant star formation history. }
\begin{tabular}{ l | l }
Parameter & Prior \\ 
 \hline \hline
 Star formation models & Bc03-MILES \\ 
 Dust law & Calzetti \\  
 Dust extinction $A_V$ & [0,2] \\
 Dust factor $\epsilon$ & [1, 10] \\
 Ionization parameter $log_{10}(U)$ & -2 \\
 Redshift & [7.6, 7.8] \\
 Mass formed $log(M^{*}/M_{\odot})$ & [7, 11] \\
 Metallicity $Z/Z_{\odot}$ & [0,2] \\
 Const. age minimum (Gyr) & [0,2] \\ 
 Const. age maximum (Gyr) & [0,2] \\  
\\  
 &
\end{tabular}
\label{tab:allpriors}
\end{table}

\begin{figure*}
    \centering
   \includegraphics[width=\textwidth]{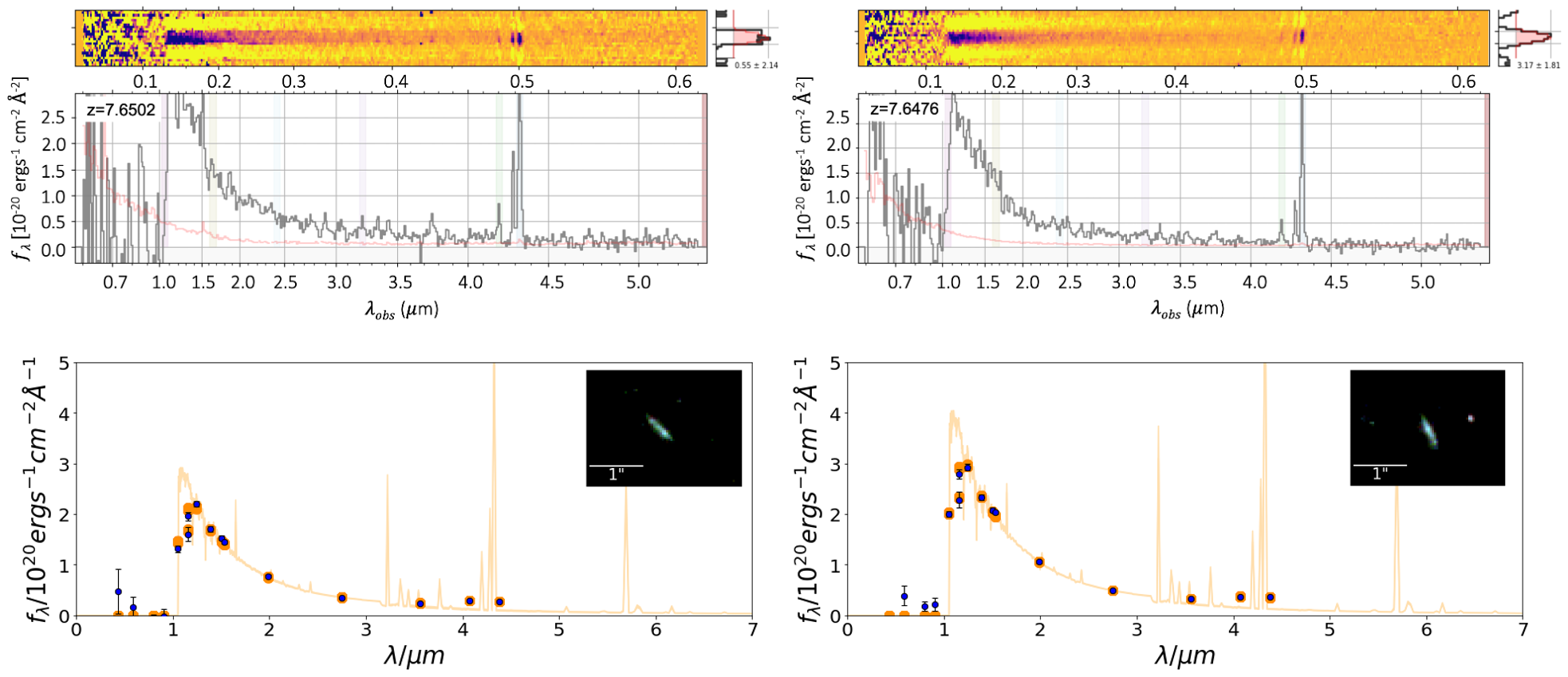}
    \caption{SED fitting and spectral analysis of the z=7.6 multiply imaged system, System 11. Top: NIRSpec 2D and 1D spectra of the two System 11 images. Bottom: The SEDs of the System 11 images fit by BAGPIPES.}
    \label{fig:sys11spec}
\end{figure*}

Adjusting for the magnification factors of the 2 images of System 11, $\mu_{11.1}=7.2^{+0.2}_{-1.2}$ and $\mu_{11.2}=8.7^{+0.4}_{-0.4}$, we use SED fitting with BAGPIPES to estimate the average stellar mass and star formation rate of this galaxy, and find $log(M/M_{\odot})$=7.37$^{+0.03}_{-0.04}$ and 3.2$^{+0.4}_{-0.2}$ $M_{\odot}/yr$ respectively. We also find an average absolute UV magnitude of -21.4$^{+0.2}_{-0.2}$, or a luminosity of $log(L/L_{\odot}) = 10.95^{+0.08}_{-0.09}$. This approximately corresponds to the knee of the UV luminosity function at $z\sim8$ \citep{Bouwens2022}. Parameters for the individual images are listed in Table \ref{tab:const_post}. The parameters derived for the two images are consistent with each other after correcting for magnification, supporting the reliability of our magnification map at high-$z$. The NIRSpec spectra and SED fits to photometry are shown in Fig. \ref{fig:sys11spec}.

\begin{table}[h]
\centering
\caption{Posterior parameters of the model with constant star formation history. The mass, star formation rate, and intrinsic luminosity as expressed in the UV magnitude have been scaled by the magnification. $M_{UV}$ was calculated from the flux in the F125W filter, which mostly closely matches the rest UV wavelength range at the galaxy's redshift. Note that the star formation history is a simple form and that therefore the age of the galaxy may be underestimated.}
\begin{tabular}{ l | l  l}
Parameter & \multicolumn{2}{c}{Posterior} \\
& Image 1 & Image 2 \\
 \hline
 Mass-weighted age (Myr) & $6^{+2}_{-2}$ & $5^{+0.001}_{-0.001}$\\ 
 
 Mass formed $log(M^{*}/M_{\odot})$ & $7.35^{+0.04}_{-0.05}$ & $7.39^{+0.02}_{-0.03}$ \\ 
Star formation rate $M_{\odot}/yr$ & $2.9_{-0.3}^{+0.4}$ & $3.6_{-0.2}^{+0.5}$ \\

 Dust extinction $A_V$ & $0.06^{+0.04}_{-0.02}$ & $0.07^{+0.08}_{-0.04}$ \\
 Dust factor $\eta$ & $3.5^{+2.2}_{-1.4}$ & $3.1^{+3.7}_{-1.6}$\\
 
 Redshift & $7.660^{+0.004}_{-0.005}$ & $7.662^{+0.003}_{-0.003}$ \\
 UV absolute magnitude $M_{UV}$ & -21.29$_{-0.19}^{+0.23}$ & -21.47$_{-0.09}^{+0.10}$
\end{tabular}
\label{tab:const_post}
\end{table}

\section{Discussion and Conclusions} \label{sec:disc}

In this paper, we have constructed a new lens model using data from JWST's NIRCam, NIRSpec, and NIRISS. The new data enabled us to re-assess the catalogue of multiple image systems and establish a subset of 24 systems with the highest confidence and spectroscopic redshifts. We removed 7 systems that were JWST-dark from the model and found spectroscopic redshifts for 3 systems.

\begin{figure*}
    \centering
        \includegraphics[width=0.8\textwidth]{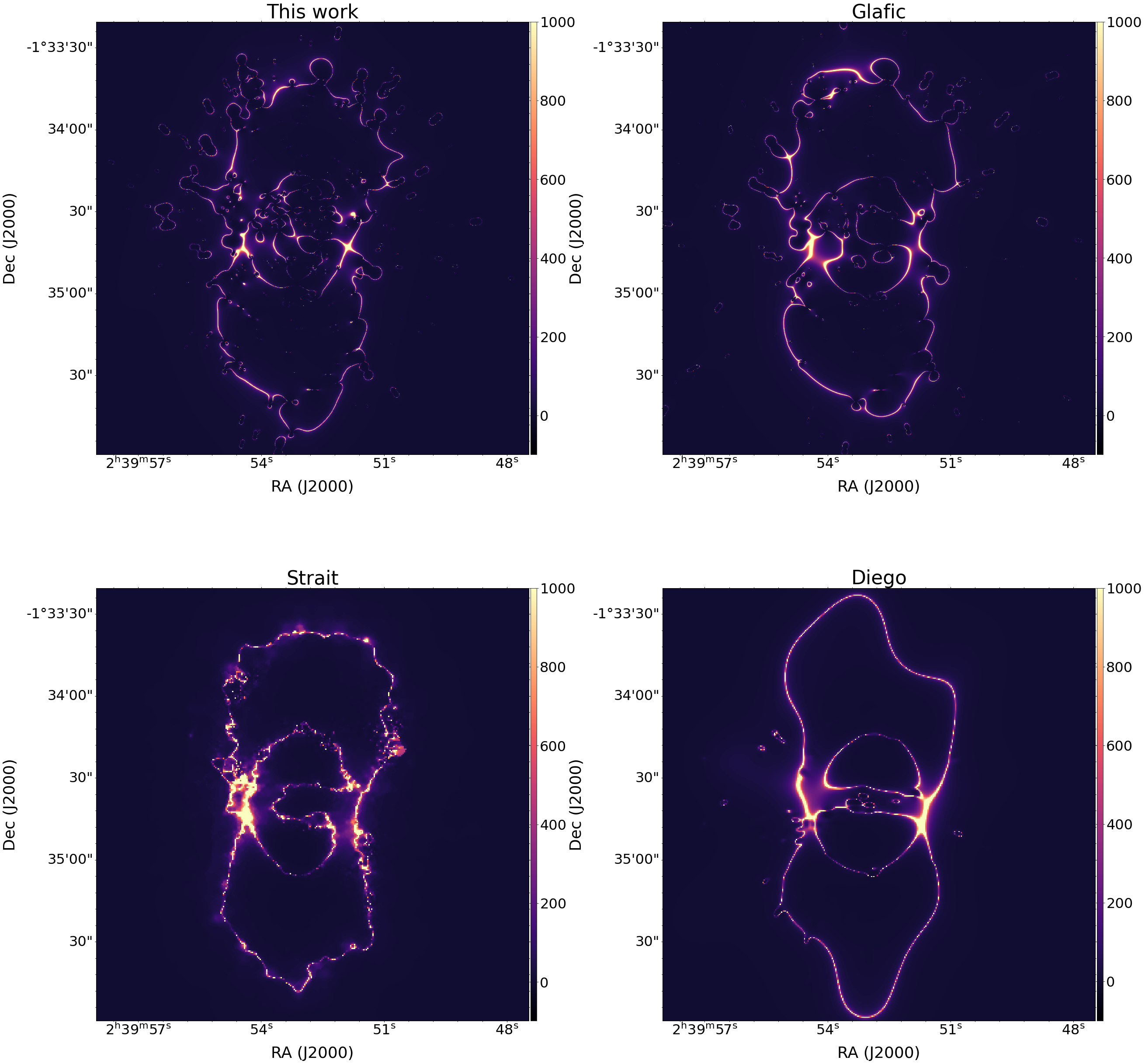}
    \caption{Magnification maps showing the critical curve for a source at z=9.0 from 4 different models of Abell 370. This work and the Glafic \citep{Kawamata2016} model are parametric while the \textcite{Strait2018} and \textcite{Diego2018} models are non-parametric. }
    \label{fig:modelcomparison}
\end{figure*}

With a few exceptions, the model accurately predicts the positions of the multiple image systems, leading to a small improvement in fit when compared to previous models. This lens model has approximately the same shape as other models constructed with different methods. Fig. \ref{fig:modelcomparison} shows some examples. Note that the \textcite{Diego2018} and \textcite{Strait2018} models are non-parametric. The latter also does not assume that light traces mass and does not include mass at the positions of the cluster members. Our model, and corresponding magnification, shear, and convergence maps are available on request and will be made publicly available on MAST in a CANUCS data release (DOI: 10.17909/ph4n-6n76).

Possible improvements of the model include providing shape information for the cluster member galaxies and increasing the computational power used to find the best-fit model.

We have also spectroscopically confirmed the redshift of a multiply-imaged galaxy at $z = 7.6476\pm0.0006$ with the characteristic luminosity at its redshift. We constrained the properties of its stellar population with SED fitting and the agreement in parameters between the two images increases confidence in the magnification map. 


\begin{acknowledgments}
This research was enabled by grant 18JWST-GTO1 from the Canadian Space Agency and funding from the Natural Sciences and Engineering Research Council of Canada.
GR and MB acknowledge support from the ERC Grant FIRSTLIGHT and from the Slovenian national research agency ARRS through grants N1-0238 and P1-0188. MB acknowledges support from the program HST-GO-16667, provided through a grant from the STScI under NASA contract NAS5-26555. This research used the Canadian Advanced Network For Astronomy Research (CANFAR) operated in partnership by the Canadian Astronomy Data Centre and The Digital Research Alliance of Canada with support from the National Research Council of Canada the Canadian Space Agency, CANARIE and the Canadian Foundation for Innovation. The Cosmic Dawn Center (DAWN) is funded by the Danish National Research Foundation under grant No. 140.
\end{acknowledgments}

%

\vspace{5mm}
\facilities{JWST (NIRSpec, NIRCam, NIRISS), HST (WFC3, ACS)}






\bibliography{library}{}
\bibliographystyle{aasjournal}

\appendix

\section{Multiple Image Catalogue}
\label{app:mi}

\begin{longtable}{cccccc}
\caption{All multiple image systems in the Abell 370 field with their positions, redshifts, and in which catalogue they were included. Systems with sub-IDs such as a, b, etc. denote knots in a system. If the system was in the Gold catalogue and therefore included in the model, the image-plane rms deviation is also provided. Most spec-z come from MUSE detections \citep{Lagattuta2019}, while the redshift of Systems 8, 11, and 43 are from NIRSpec. The number of significant figures reflects the uncertainty in the MUSE redshifts. } \\
\textbf{ID} & \textbf{RA} & \textbf{Dec} & \textbf{Redshift} & \textbf{RMS ($\arcsec$)} & \textbf{Catalogue}\\ 
\hline \hline
1a.1 & 39.9762942 & -1.5760362 & 0.8041 & 0.673 & Gold \\
1a.2 & 39.9671518 & -1.576882 & 0.8041 & 0.471 & Gold \\
1a.3 & 39.9685746 & -1.5766236 & 0.8041 & 0.230 & Gold \\
1b.2 & 39.9675631 & -1.5767597 & 0.8041 & 0.06 & Gold \\
1b.3 & 39.9680966 & -1.5766598 & 0.8041 & 0.06 & Gold \\
1c.2 & 39.9670351 & -1.5769042 & 0.8041 & 0.207 & Gold \\
1c.3 & 39.9687358 & -1.5765875 & 0.8041 & 0.207 & Gold \\
\hline
2.1 & 39.973789 & -1.584241 & 0.7251 & 0.496 & Gold \\
2.2 & 39.970973 & -1.585035 & 0.7251 & 0.207 & Gold \\
2.3 & 39.968741 & -1.584507 & 0.7251 & 0.557 & Gold \\
\hline
3a.1 & 39.9685154 & -1.5658046 & 1.9553 & 0.573 & Gold \\
3a.2 & 39.965677 & -1.5668566 & 1.9553 & 0.331 & Gold \\
3a.3 & 39.9789537 & -1.5674604 & 1.9553 & 0.543 & Gold \\
3b.1 & 39.9673222 & -1.5661306 & 1.9553 & 0.534 & Gold \\
3b.2 & 39.9661958 & -1.5665973 & 1.9553 & 0.161 & Gold \\
3b.3 & 39.9790446 & -1.567469 & 1.9553 & 0.663 & Gold \\
\hline
4a.1 & 39.9796721 & -1.5763896 & 1.2728 & 0.617 & Gold \\
4a.2 & 39.9707234 & -1.5762784 & 1.2728 & 0.547 & Gold \\
4a.3 & 39.9619525 & -1.5779415 & 1.2728 & 0.487 & Gold \\
4b.1 & 39.9797684 & -1.5765674 & 1.2728 & 0.657 & Gold \\
4b.2 & 39.9706752 & -1.5764562 & 1.2728 & 0.492 & Gold \\
4b.3 & 39.9620043 & -1.5780896 & 1.2728 & 0.457 & Gold \\
4c.1 & 39.9795892 & -1.5762564 & 1.2728 & 0.551 & Gold \\
4c.2 & 39.9707553 & -1.5761507 & 1.2728 & 0.574 & Gold \\
4c.3 & 39.9619075 & -1.5778271 & 1.2728 & 0.535 & Gold \\
\hline
5.1 & 39.973486 & -1.58905 & 1.2775 & 0.478 & Gold \\
5.2 & 39.971018 & -1.589217 & 1.2775 & 0.263 & Gold \\
5.3 & 39.96913 & -1.589053 & 1.2775 & 0.368 & Gold \\
\hline
6a.1 & 39.9797577 & -1.5772207 & 1.0633 & 0.252 & Gold \\
6a.2 & 39.9693676 & -1.577354 & 1.0633 & 0.155 & Gold \\
6a.3 & 39.9644985 & -1.5783614 & 1.0633 & 0.348 & Gold \\
6b.1 & 39.9796515 & -1.5770878 & 1.0633 & 0.372 & Gold \\
6b.2 & 39.969486 & -1.5771825 & 1.0633 & 0.15 & Gold \\
6b.3 & 39.9643411 & -1.5782225 & 1.0633 & 0.343 & Gold \\
6c.1 & 39.9794544 & -1.5770341 & 1.0633 & 0.367 & Gold \\
6c.2 & 39.969659 & -1.5770966 & 1.0633 & 0.182 & Gold \\
6c.3 & 39.9640847 & -1.5781854 & 1.0633 & 0.428 & Gold \\
\hline
7.1 & 39.9698256 & -1.580573 & 2.7512 & 0.638 & Gold \\
7.3 & 39.968808 & -1.5856333 & 2.7512 & ----- & Silver \\
7.4 & 39.986554 & -1.5775806 & 2.7512 & ----- & Silver \\
7.5 & 39.961542 & -1.5800056 & 2.7512 & ----- & Silver \\
7.6 & 39.9683489 & -1.5713448 & 2.7512 & 0.638 & Gold \\
\hline
8.1 & 39.9645508 & -1.5697533 & 2.866 & 0.06 & Gold \\
8.2 & 39.961868 & -1.5736833 & 2.866 & 0.06 & Gold \\
\hline
9.1 & 39.9624 & -1.5778861 & 1.5182 & 0.662 & Gold \\
9.2 & 39.969483 & -1.5762667 & 1.5182 & 0.629 & Gold \\
9.3 & 39.982017 & -1.5765333 & 1.5182 & 0.15 & Gold \\
\hline
11a.1 & 39.9637986 & -1.5693826 & 7.6476 & 0.099 & Gold \\
11a.2 & 39.960764 & -1.5741493 & 7.6476 & 0.099 & Gold \\
11b.1 & 39.9638653 & -1.569316 & 7.6476 & 0.127 & Gold \\
11b.2 & 39.9607279 & -1.5742159 & 7.6476 & 0.127 & Gold \\
\hline
12a.1 & 39.9841376 & -1.5708867 & 3.4809 & 0.736 & Gold \\
12a.2 & 39.9696122 & -1.5666344 & 3.4809 & 0.302 & Gold \\
12a.3 & 39.9592413 & -1.5752348 & 3.4809 & 0.552 & Gold \\
12b.1 & 39.9840783 & -1.5709608 & 3.4809 & 0.755 & Gold \\
12b.2 & 39.9697827 & -1.5666566 & 3.4809 & 0.236 & Gold \\
12b.3 & 39.9591524 & -1.5754052 & 3.4809 & 0.620 & Gold \\
12c.1 & 39.9839791 & -1.5707975 & 3.4809 & 0.672 & Gold \\
12c.2 & 39.9698687 & -1.5665785 & 3.4809 & 0.292 & Gold \\
12c.3 & 39.9591413 & -1.5752015 & 3.4809 & 0.501 & Gold \\
\hline
13a.1 & 39.9795197 & -1.5717758 & 4.248 & 0.616 & Gold \\
13a.2 & 39.9752181 & -1.5688202 & 4.248 & 0.947 & Gold \\
13a.3 & 39.9567562 & -1.5774992 & 4.248 & 1.426 & Gold \\
13b.1 & 39.9795809 & -1.5718161 & 4.248 & 0.544 & Gold \\
13b.2 & 39.9751362 & -1.5687702 & 4.248 & 0.544 & Gold \\
\hline
14a.1 & 39.9813269 & -1.5781642 & 3.1309 & 0.489 & Gold \\
14a.2 & 39.9722981 & -1.5780015 & 3.1309 & 0.771 & Gold \\
14a.3 & 39.9721943 & -1.5800387 & 3.1309 & 1.384 & Gold \\
14a.4 & 39.9576976 & -1.5804348 & 3.1309 & 0.401 & Gold \\
14a.5 & 39.9742092 & -1.5856151 & 3.1309 & 0.638 & Gold \\
14b.1 & 39.9812899 & -1.578279 & 3.1309 & 0.207 & Gold \\
14b.2 & 39.9723129 & -1.5782385 & 3.1309 & 0.461 & Gold \\
14b.4 & 39.957679 & -1.5805015 & 3.1309 & 0.772 & Gold \\
14b.5 & 39.9743259 & -1.5855799 & 3.1309 & 0.289 & Gold \\
\hline
15.1 & 39.971328 & -1.580604 & 3.7085 & ----- & Bronze \\
15.2 & 39.971935 & -1.5870512 & 3.7085 & ----- & Bronze \\
15.3 & 39.971027 & -1.5777907 & 3.7085 & ----- & Bronze \\
15.4 & 39.984017 & -1.5784514 & 3.7085 & ----- & Bronze \\
\hline
16.1 & 39.964016 & -1.5880782 & 3.7743 & ----- & Silver \\
16.2 & 39.966037 & -1.5890355 & 3.7743 & ----- & Bronze \\
16.3 & 39.984414 & -1.5841111 & 3.7743 & ----- & Silver \\
\hline
17.1 & 39.969758 & -1.5885333 & 4.2567 & 0.446 & Gold \\
17.2 & 39.985403 & -1.5808406 & 4.2567 & 0.562 & Gold \\
17.3 & 39.960235 & -1.5836508 & 4.2567 & 0.366 & Gold \\
\hline
18.1 & 39.97583 & -1.5870613 & 4.4296 & 0.234 & Gold \\
18.2 & 39.981476 & -1.5820728 & 4.4296 & 0.234 & Gold \\
\hline
19.1 & 39.971996 & -1.5878654 & 5.6493 & ----- & Silver \\
19.2 & 39.985142 & -1.5790944 & 5.6493 & ----- & Silver \\
19.3 & 39.958316 & -1.5813093 & 5.6493 & ----- & Silver \\
\hline
20.1 & 39.965271 & -1.5878028 & 5.7505 & 0.555 & Gold \\
20.2 & 39.963608 & -1.5868833 & 5.7505 & 0.555 & Gold \\
\hline
21.1 & 39.966575 & -1.5846139 & 1.2567 & ----- & Bronze \\
21.2 & 39.967383 & -1.5850278 & 1.2567 & ----- & Bronze \\
21.3 & 39.981539 & -1.5814028 & 1.2567 & ----- & Bronze \\
\hline
22a.1 & 39.981683 & -1.5796827 & 3.1309 & 0.328 & Gold \\
22a.2 & 39.9744122 & -1.5861063 & 3.1309 & 0.328 & Gold \\
22b.1 & 39.9816163 & -1.5798189 & 3.1309 & 0.307 & Gold \\
22b.2 & 39.9745178 & -1.5860563 & 3.1309 & 0.307 & Gold \\
\hline
23.1 & 39.980254 & -1.5667639 & 5.9386 & ----- & Silver \\
23.2 & 39.957314 & -1.572744 & 5.9386 & ----- & Silver \\
23.3 & 39.977165 & -1.5662748 & 5.9386 & ----- & Silver \\
\hline
24a.1 & 39.9635133 & -1.5702107 & 4.9153 & 0.051 & Gold \\
24a.2 & 39.9615735 & -1.5734148 & 4.9153 & 0.051 & Gold \\
24b.1 & 39.9634818 & -1.5702422 & 4.9153 & 0.051 & Gold \\
24b.2 & 39.9615957 & -1.573374 & 4.9153 & 0.051 & Gold \\
24c.1 & 39.9632836 & -1.5703459 & 4.9153 & 0.067 & Gold \\
24c.2 & 39.9617365 & -1.572911 & 4.9153 & 0.067 & Gold \\
24d.1 & 39.963114 & -1.5706099 & 4.9153 & 0.268 & Gold \\
24d.2 & 39.9620561 & -1.5723396 & 4.9153 & 0.268 & Gold \\
24e.1 & 39.9630029 & -1.5706265 & 4.916 & 0.249 & Gold \\
24e.2 & 39.9621432 & -1.5719822 & 4.916 & 0.249 & Gold \\
24f.1 & 39.9630844 & -1.570708 & 4.916 & 0.352 & Gold \\
24f.2 & 39.9622377 & -1.5720674 & 4.916 & 0.352 & Gold \\
\hline
25a.1 & 39.9872963 & -1.5787605 & 3.8145 & 0.099 & Gold \\
25a.2 & 39.9617351 & -1.5828757 & 3.8145 & 0.099 & Gold \\
25b.1 & 39.9872393 & -1.5789939 & 3.8145 & 0.137 & Gold \\
25b.2 & 39.9619074 & -1.5832451 & 3.8145 & 0.137 & Gold \\
25c.1 & 39.987274 & -1.5788161 & 3.8145 & 0.103 & Gold \\
25c.2 & 39.9617546 & -1.5829368 & 3.8145 & 0.103 & Gold \\
\hline
26.1 & 39.979939 & -1.5713902 & 3.9359 & 0.541 & Gold \\
26.2 & 39.974464 & -1.5680938 & 3.9359 & 0.541 & Gold \\
26.3 & 39.957165 & -1.5769585 & 3.9359 & ----- & Silver \\
\hline
27.1 & 39.972446 & -1.567157 & 3.0161 & 0.492 & Gold \\
27.2 & 39.980694 & -1.571125 & 3.0161 & 0.372 & Gold \\
27.3 & 39.95829 & -1.5759068 & 3.0161 & 0.15 & Gold \\
\hline
28.1 & 39.963492 & -1.5822806 & 2.9101 & 0.665 & Gold \\
28.2 & 39.967058 & -1.5845583 & 2.9101 & 0.665 & Gold \\
28.3 & 39.987816 & -1.5774528 & 2.9101 & ----- & Silver \\
\hline
29a.1 & 39.9834878 & -1.5673115 & 4.4897 & 0.487 & Gold \\
29a.2 & 39.9684473 & -1.5646408 & 4.4897 & 0.476 & Gold \\
29a.3 & 39.9601989 & -1.5694708 & 4.4897 & 0.837 & Gold \\
29b.1 & 39.9835804 & -1.5675745 & 4.4897 & 0.991 & Gold \\
29b.2 & 39.9676432 & -1.5649075 & 4.4897 & 0.079 & Gold \\
29b.3 & 39.9609326 & -1.5690004 & 4.4897 & 0.952 & Gold \\
\hline
30.1 & 39.983351 & -1.5704081 & 5.6459 & ----- & Silver \\
30.2 & 39.972404 & -1.5663533 & 5.6459 & ----- & Silver \\
\hline
31.1 & 39.972404 & -1.5693301 & 5.4476 & ----- & Quartz \\
31.2 & 39.980667 & -1.5747346 & 5.4476 & ----- & Quartz \\
31.3 & 39.956158 & -1.5786786 & 5.4476 & ----- & Quartz \\
\hline
32.1 & 39.966286 & -1.5693446 & 4.4953 & ----- & Bronze \\
32.2 & 39.988098 & -1.5751871 & 4.4953 & ----- & Bronze \\
32.3 & 39.960682 & -1.5783795 & 4.4953 & ----- & Bronze \\
\hline
33.1 & 39.962723 & -1.5860035 & 4.882 & ----- & Quartz \\
33.2 & 39.966217 & -1.5879961 & 4.882 & ----- & Quartz \\
\hline
34.1 & 39.970108 & -1.5701499 & 5.2437 & ----- & Quartz \\
34.2 & 39.971806 & -1.5880395 & 5.2437 & ----- & Quartz \\
34.3 & 39.958565 & -1.5817008 & 5.2437 & ----- & Quartz \\
34.4 & 39.985046 & -1.579559 & 5.2437 & ----- & Quartz \\
\hline
35.1 & 39.981541 & -1.5658624 & 6.1735 & ----- & Quartz \\
35.2 & 39.975826 & -1.5644423 & 6.1735 & ----- & Quartz \\
\hline
36.1 & 39.962444 & -1.5807098 & 6.2855 & ----- & Quartz \\
36.2 & 39.965996 & -1.5843844 & 6.2855 & ----- & Quartz \\
\hline
37.1 & 39.97039 & -1.5687943 & 5.6489 & ----- & Quartz \\
37.2 & 39.970428 & -1.5694203 & 5.6489 & ----- & Quartz \\
\hline
38.1 & 39.9771985 & -1.5738047 & 3.2 & 0.684 & Gold \\
38.2 & 39.9750827 & -1.5721194 & 3.2 & 0.684 & Gold \\
\hline
39.1 & 39.965442 & -1.5780222 & 1.2777 & ----- & Bronze \\
39.2 & 39.967933 & -1.5773472 & 1.2777 & ----- & Bronze \\
39.3 & 39.982296 & -1.576975 & 1.2777 & ----- & Bronze \\
\hline
40.1 & 39.963579 & -1.5656333 & 1.0323 & ----- & Bronze \\
40.2 & 39.962958 & -1.5661111 & 1.0323 & ----- & Bronze \\
40.3 & 39.963375 & -1.5659528 & 1.0323 & ----- & Bronze \\
\hline
41.1 & 39.970546 & -1.5693801 & 4.9441 & ----- & Quartz \\
41.2 & 39.969977 & -1.5700367 & 4.9441 & ----- & Quartz \\
41.3 & 39.985223 & -1.5793885 & 4.9441 & ----- & Quartz \\
41.4 & 39.971395 & -1.58802 & 4.9441 & ----- & Quartz \\
\hline
42.1 & 39.970632 & -1.5710393 & 4.3381 & ----- & Bronze \\
42.2 & 39.983162 & -1.5796664 & 4.3381 & ----- & Bronze \\
42.3 & 39.973383 & -1.5874465 & 4.3381 & ----- & Bronze \\
42.4 & 39.957967 & -1.5815081 & 4.3381 & ----- & Bronze \\
\hline
43.1 & 39.9824 & -1.5811 & 8.1970 & ----- & Silver \\
43.2 & 39.975875 & -1.58726 & 8.1970 & ----- & Silver \\
\label{tab:allMI}
\end{longtable}

\end{document}